# Inelastic neutron scattering investigations of an anisotropic hybridization gap in the Kondo insulators: $CeT_2Al_{10}$ (T=Fe, Ru and Os)

D.T. Adroja[1,2,4] [*], Y. Muro[3], T. Takabatake[,4], M.D. Le[1], H. C. Walker[1], K.A. McEwen[5] and A.T. Boothroyd[6]

[1]ISIS Facility, Rutherford Appleton Laboratory, Chilton, Didcot, Oxon, OX11 0QX, UK

[2]Highly correlated electron group, Physics Department, University of Johannesburg, PO Box 524, Auckland Park 2006, South Africa

[3]Faculty of Engineering, Toyama Prefectural University, Imizu 939-0398, Japan

[4]Department of Quantum Matter, ADSM, and IAMR, Hiroshima University, Higashi-Hiroshima, 739-8530, Japan

[5]Department of Physics and Astronomy, and London Centre for Nanotechnology,

University College London, Gower Street, London WC1E 6BT, UK

[6] Department of Physics, University of Oxford, Clarendon Laboratory, Parks Road, Oxford OX1 3PU, United Kingdom

[*]devashibhai.adroja@stfc.ac.uk

**Abstract.** The recent discovery of topological Kondo insulating behaviour in strongly correlated electron systems has generated considerable interest in Kondo insulators both experimentally and theoretically. The Kondo semiconductors $CeT_2Al_{10}$ (T=Fe, Ru and Os) possessing a *c-f* hybridization gap have received considerable attention recently because of the unexpected high magnetic ordering temperature of $CeRu_2Al_{10}$ ($T_N$=27 K) and $CeOs_2Al_{10}$ ($T_N$=28.5 K) and the Kondo insulating behaviour observed in the valence fluctuating compound $CeFe_2Al_{10}$ with a paramagnetic ground state down to 50 mK. We are investigating this family of compounds, both in polycrystalline and single crystal form, using inelastic neutron scattering to understand the role of anisotropic *c-f* hybridization on the spin gap formation as well as on their magnetic properties. We have observed a clear sign of a spin gap in all three compounds from our polycrystalline study as well as the existence of a spin gap above the magnetic ordering temperature in T=Ru and Os. Our inelastic neutron scattering studies on single crystals of $CeRu_2Al_{10}$ and $CeOs_2Al_{10}$ revealed dispersive gapped spin wave excitations below $T_N$. Analysis of the spin wave spectrum reveals the presence of strong anisotropic exchange, along the c-axis (or z-axis) stronger than in the ab-plane. These anisotropic exchange interactions force the magnetic moment to align along the c-axis, competing with the single ion crystal field anisotropy, which prefers moments along the a-axis. In the paramagnetic state (below 50 K) of the Kondo insulator $CeFe_2Al_{10}$, we have also observed dispersive gapped magnetic excitations which transform into quasi-elastic scattering on heating to 100 K. We will discuss the origin of the anisotropic hybridization gap in $CeFe_2Al_{10}$ based on theoretical models of heavy-fermion semiconductors.

# 1. Introduction

The investigation of the effect of strong electronic correlations on the physical properties of d- and f-electron systems is an exciting and diverse field in both experimental and theoretical condensed matter physics. The discovery of heavy fermion compounds in 1975, high-temperature superconductors in 1986, and topological Kondo insulators in 2012, has generated interest in the field of strongly correlated electron systems. These are systems in which the strength of the electron-electron interactions is comparable to, or larger than, the kinetic energy of the electrons. Among the strongly correlated electron systems, Ce-, Yb- and U-based strongly correlated electron systems with f-electrons have attracted considerable interest due to the duality between the itinerant and the localized nature of f-electrons that gives rise to a rich variety of novel phenomena. This includes, for example, heavy electron/fermion behavior [1], mixed valence/valence fluctuations behavior [2], reduced magnetic moment ordering [3], anomalously high magnetic ordering temperatures compared with their isostructural Gd-based compounds [4], unconventional superconductivity [5], Kondo insulators or Kondo semiconductors [6,7], spin and charge gap formation, metal-insulator transition [8], non-Fermi-liquid (NFL) behavior and zero-temperature quantum critical phase transitions [9,10,11].

Many of above mentioned phenomena arise due to the presence of strong electron-electron correlations as revealed by a large enhancement of low-temperature properties, such as the observed huge electronic heat capacity coefficient, up to 1000 times or more than that observed in normal metals, and an enhanced value of the low temperature susceptibility above the values expected from local density approximation electronic-structure calculations. The first heavy fermion behavior was discovered in $CeAl_3$ by Andres, Graebner and Ott in 1975 [12], who observed enormous magnitudes of the linear specific heat capacity, $\gamma_{ele}$=1620 mJ/mol-K$^2$. On the other hand, existence of localized magnetic moments and superconductivity was observed in the heavy fermion superconductor $CeCu_2Si_2$ by Steglich et al in 1979 [5], while Kondo insulating behavior with a small energy gap ~4-14meV in $SmB_6$ was reported by Allen, Batlogg and Wachter in 1979 [13]. Unlike conventional band insulators, a Kondo insulator (KI) features an energy gap in the electronic density of states (DOS) near the Fermi level ($E_F$) whose magnitude is strongly temperature dependent and the full gap only develops at low temperatures.

The different physical properties, local moment, heavy fermion, Kondo effect and valence fluctuations discussed above can be understood using a variable degree of hybridization ($V_{cf}$) between localized *f* (or *d*) electrons and conduction electrons, *c*, including onsite Coulomb interactions U, first proposed by P. Anderson in 1961 for a magnetic impurity known as the Anderson impurity model (AIM) [14]. In the weak coupling limit (i.e. small value of $V_{cf}$) this model gives rise to spin fluctuations, while in the strong coupling limit both spin and charge fluctuations are observed. To explain the effect of a magnetic impurity on the low-temperature resistivity ($\rho(T)\sim-\ln(T)$) Jun Kondo in 1964 [15] proposed a model based on an onsite antiferromagnetic interaction ($J_{cf}$) between the local moment and conduction electrons (Fig.1a). In 1966 Schrieffer and Wolff [16] proposed a canonical transformation, treating the hybridization term in perturbation theory up to second order in $V_{cf}$ that provided a direct relation between the Anderson and Kondo models. The physics of strongly correlated systems is governed by a direct competition between two interactions. One is the onsite Kondo interaction ($T_K$) of the local moment with conduction electrons that screens the local moment by forming a non-magnetic Kondo singlet, resulting in a non-magnetic singlet state. The other is inter-site Ruderman-Kittel-Kasuya-Yosida (RKKY, $T_{RKKY}$) interaction (1954-1957 [17]), which stabilizes the magnetic ground state. Tuning the localized *f*- (or *d*-) electron's energy ($E_f$) toward the Fermi level ($E_F$) increases the strength of coupling ($V_{cf}$ or $J_{cf}$) and leads a crossover between the Kondo regime and mixed valence regime of the Anderson model (Fig. 2).

Although the AIM and Kondo model were successful in explaining many observed properties of strongly correlated electron systems, they failed to explain some of the observed properties of a dense Kondo system (Kondo lattice), where Kondo centres are arranged in the form of a lattice (Fig.1b), and exhibit an onset of coherence below a characteristic temperature, T*. Below T*, strong elastic scattering disappears and individual Kondo centres coherently scatter conduction electrons and hence the local 4$f$ (or 5$f$) moments become partially itinerant, giving rise to the universal Kondo liquid behaviour and rich phase diagrams. The simplest model to describe the behaviour of dense Kondo lattices and Kondo insulators is the periodic Anderson model (PAM) [18]. The PAM is exactly solvable in the non-interacting limit when U → ∞. Riseborough [6,19] has obtained the solution for the PAM with orbital degeneracy N = 2, in the mean-field slave-boson approximation, which reveals the two-band pictures, upper and lower hybridized bands separated by a hybridization gap. At half filling this exhibits a direct gap (which we denote as the charge gap, $\Delta_{char}$ or $\Delta_{dir}$) at q = 0 and an indirect gap (spin gap, $\Delta_{spin}$ or $\Delta_{ind}$) at q ≠ 0 in the fermionic density of states near $E_F$ (Fig.1c). The magnitudes of the indirect and direct gaps are given by $\Delta_{ind}=2V_{cf}^2/W$ and $\Delta_{dir}=2V_{cf}= (WT_K)^{1/2}$, respectively, where W is the half width of the conduction band and $V_{cf}$ is the hybridization matrix element. From the mean field solution of the Kondo impurity problem, we can anticipate $V_{cf}^2/W \sim V_{cf}^2\rho(E_F) \sim T_K$ and hence $\Delta_{ind}$ (or $\Delta_{spin}$) ~ $T_K$ [18]. The gap widths inferred from optical, magnetic, transport, and thermodynamics properties of Kondo insulators are almost an order of magnitude smaller than those obtained by band structure calculations, which are due to the presence of strong electronic correlations. As a consequence of strong c-f hybridization and the formation of a Kondo singlet ground state, the Kondo insulating ground state is not compatible with magnetic ordering. For example, Kondo insulators CeNiSn [20], CeRhSb [21], $Ce_3Pt_3Bi_4$ [22], $SmB_6$ [13, 23] and $YbB_{12}$ [24] remain paramagnetic and no long range magnetic ordering has been observed in these systems. Theoretical models were proposed to explain the anisotropic properties of CeNiSn by Ikeda and Miyake (IM) based on anisotropic hybridization in k-space [25], and by Moreno and Coleman (MC) based on global and local minima in the crystal field (CEF) parameter space to explain a gap anisotropy [26].

The Kondo temperature varies exponentially with the product of the electron density of states $\rho(E_F)$ at $E_F$ and the exchange coupling $J_{cf}$ between conduction electrons and local moments, $T_K \sim (\rho(E_F)*J_{cf})^{1/2}*\exp(-1/|\rho(E_F)*J_{cf}|)$, while the magnetic RKKY interaction energy varies as square of this product, $T_{RKKY} \sim \rho(E_F)*(J_{cf})^2$. The value of $\rho(E_F)$ can be estimated using the low-temperature heat capacity, $\rho(E_F)=3*\gamma_{ele}/\pi^2$, where $\gamma_{ele}$ is Sommerfeld coefficient. Doniach [27] made the first theoretical study of a Kondo lattice using a one-dimensional chain of spins and this result is still used to classify many heavy fermion and strongly correlated compounds on the phase diagram (Fig. 2). In a simple picture, the magnetically ordering temperature $T_N$ initially increases with increasing $J_{cf}$, then passes through a maximum and vanishes ($T_N$ -> 0) at a critical coupling $J_{cf}^c$, where a quantum critical point (QCP) exits. Near QCP, the well known Landau Fermi-liquid picture breaks down and non-Fermi-liquid (NFL) behavior has been observed [9]. This behavior of $T_N$ and QCP/NFL has been experimentally observed in many cerium-based compounds by varying the pressure, magnetic field and concentration [1, 9, 10].

Very recently the caged type compounds $CeT_2Al_{10}$ (T=Fe, Ru and Os) have attracted considerable attention due to Kondo insulating behavior and the opening of spin and charge gaps [28]. In the present work we have investigated a spin gap formation in $CeT_2Al_{10}$ (T=Fe, Ru and Os) compounds using time-of-flight (TOF) inelastic neutron scattering measurements at the ISIS facility. We have found a clear sign of dispersive magnetic excitations in these compounds. We will discuss the origin of the magnetic excitations in these compounds using various theoretical models.

## 2. Results and discussion

### 2.1. Kondo insulating behavior of $CeT_2Al_{10}$

Recent investigations on CeT$_2$Al$_{10}$ (T=Fe, Ru and Os), which crystallise in the orthorhombic structure (space group No 63 Cmcm) (Fig. 3), have generated considerable interest in both theoretical and experimental condensed matter physics [28, 29-48]. This interest arose due to the various interesting ground states observed in this family of Ce-compounds as well as their caged type crystal structure that is important for the enhanced thermoelectric properties. The resistivity of CeRu$_2$Al$_{10}$ exhibits a sharp drop near 27 K resembling an insulator-metal transition [28, 29]. A very similar phase transition, near 29 K, has been observed in CeOs$_2$Al$_{10}$ [28, 30]. The susceptibility (along the a-axis) exhibits a broad maximum near 45 K above T$_N$=29 K in contrast to a sharp drop at the phase transition (27 K) in CeRu$_2$Al$_{10}$ [28,29]. The broad maximum in the susceptibility indicates the opening of a spin or hybridization gap (Fig.4). On the other hand, the susceptibility of CeFe$_2$Al$_{10}$ along the a-axis exhibits a broad maximum near 75 K without any sign of magnetic ordering [28, 33], which indicates that the Ce ions are in the valence fluctuating or mixed valence state, as confirmed by L$_3$-edge x-ray absorption spectroscopy [46]. Further the strong anisotropic behavior in the paramagnetic state of all three compounds reveals the presence of strong hybridization between 4$f$- and conduction electrons. Moreover, there is also a strong single ion anisotropy arising from the crystal field potential in T=Ru and Os [47, 48]. Another difference appears in the resistivity of CeT$_2$Al$_{10}$ (T=Os and Ru). The resistivity for T=Os displays a thermal activation-type temperature dependence below 15 K while the resistivity for T=Ru exhibits "metallic" behaviour below the phase transition down to 2 K (Fig. 4). More interesting is the revelation from neutron diffraction that the Ce ordered moments, with propagation vector K= [1 0 0], are very small ($\mu_{ord}$=0.32$\mu_B$ for T=Ru and $\mu_{ord}$=0.39$\mu_B$ T=Os), and furthermore are directed along the c-axis [37,39], which is not the direction preferred by the crystal field anisotropy that seeks to align the moment along the a-axis [47, 48]. Therefore, systematic investigations of CeT$_2$Al$_{10}$ (T = Fe, Ru, and Os) with different values of the Kondo temperature T$_K$ have been performed to reveal the role of the *c-f* hybridization in the mysterious phase transition and gap formation.

## 2.2 Inelastic neutron scattering

Inelastic neutron scattering (INS) is a powerful method which allows us to determine both the spatial and time correlations of the magnetic excitations of the sample through the dipolar coupling of the magnetic moment of the neutron with the spin-correlation function of the sample [50-53]. The neutron is a probe that provides a magnetic perturbation that varies in space and time (or a wave-vector and frequency dependent magnetic field) and hence allows the dynamic response of the sample to be measured. The energy of thermal neutrons is comparable to the energy scale of magnetic excitations and lattice vibrations (or phonons) in solid state materials and hence INS is ideally matched to provide vital information on magnetic exchange interactions as well as lattice dynamics. INS studies have been extensively used to understand the origin of superconductivity in Cu-based and Fe-based materials as well as unconventional superconductivity in heavy fermion materials by investigating high- and low-energy dispersive spin excitations, spin gap formation and spin resonances together with phonon dispersions [23, 24, 53, 54]. In the present systems, CeT$_2$Al$_{10}$ (T=Ru and Os), the INS technique was used to show that the single-ion CEF anisotropy is insufficient to explain the observed gapped spin wave excitations. Rather, an anisotropic exchange interaction, due to hybridization between 4$f$- and conduction-electrons, is needed to fit the data. The latter two contributions are also investigated in Ce(Ru$_{1-x}$Fe$_x$)$_2$Al$_{10}$ with x=0.8 and CeFe$_2$Al$_{10}$, which have nonmagnetic ground states down to 1.2 K and 40 mK, respectively [46].

### 2.2.1 INS study on polycrystalline CeT$_2$A$_{l0}$

The inelastic neutron scattering study (INS) on CeT$_2$Al$_{10}$ (T=Fe, Ru and Os) reveals a clear sign of the spin-gap formation of 8 meV in T=Ru [44, 46], 11 meV in T=Os [28, 38] and 12 meV in T=Fe [46, 54]. The temperature dependent study of T=Ru and Os reveals that the spin gap decreases with increasing temperature in T=Ru and Os [28, 44, 46], but a finite value of the gap still exists above T$_N$ (Fig.5). By raising the temperature still further (above 40 K), the INS response becomes very broad, of quasi-elastic-type for T=Ru and Os [28, 44, 46]. A very similar change in the excitation spectrum has been observed for $CeFe_2Al_{10}$ when the temperature was increased from 5 K to 100 K [46], which is in agreement with that reported for $CeFe_2Al_{10}$ single crystal by Mignot et al [54]. It is interesting to note that despite the spin gap opening at 45 K, $CeOs_2Al_{10}$ exhibits long-range magnetic ordering below 29 K. Thus $CeOs_2Al_{10}$ is a unique system and probably the first Ce-based system exhibiting a Kondo insulating gap and long-range magnetic ordering. All Kondo insulators known so far, such as CeNiSn, CeRhSb, $Ce_3Pt_3Bi_4$, $SmB_6$, $YbB_6$ and FeSi, are paramagnetic and do not exhibit long-range magnetic ordering down to the lowest measured temperature [20-24]. Furthermore in the proposed phase diagram of Fig.2, the majority of Kondo insulators are in the non-magnetic heavy fermion and mixed valence regimes of the phase diagram, but $CeT_2Al_{10}$ (T=Ru and Os) are on the magnetic side. Therefore, a new theoretical approach is required to explain the observed coexistence of the Kondo insulating and long-range ordered states.

It is an open question, which interactions control the magnitude of spin gap or hybridization gap in Kondo insulators, $T_K$, $T^*$, $T_{RKKY}$ or $V_{cf}$. According to the single impurity model [55, 56], we can estimate the high temperature Kondo temperature $T_K$ through the maximum $T^{\chi}_{max}$ in the bulk magnetic susceptibility as $T_K= 3* T^{\chi}_{max}$. It is interesting to note that for many strongly correlated electron systems including Kondo insulators, a universal scaling relation was observed between the spin gap energy (i.e. peak position observed in the INS study) and the Kondo temperature ($T_K=3*T^{\chi}_{max}$, where $T^{\chi}_{max}$ is the temperature where susceptibility exhibits a maximum or a peak) [38, 56, 57]. For example, in the mixed valence Kondo insulator $Ce_3Pt_3Bi_4$, the energy of a peak near 20 meV (an onset near 12meV) in the INS spectrum corresponds to $T_K= 3* T^{\chi}_{max\,K} =180$ K=15.5 meV which is estimated from the peak in the susceptibility at 60 K [58]. Further, the Kondo insulator CeNiSn exhibits a spin gap in INS studies at 2 meV (along (0 0 L) and (0 K 0) directions) and 4 meV (along ($Q_a$, ½+n, $Q_c$) [59], where n is an integer and $Q_a$ and $Q_c$ are arbitrary), so the average value of the gap is 3 meV. This energy agrees with $T_K$=36 K=3.1 meV derived from the susceptibility maximum temperature at 12 K along the a-axis [20, 59]. On the other hand, the Kondo semimetal $CeRu_4Sb_{12}$ exhibits a peak in the INS study near 30 meV, while $T_K=3*100K=300K=26$ meV was estimated from the susceptibility maximum at 100K [60].

Now if we apply a similar scaling idea to the spin gap in our polycrystalline study of $CeT_2Al_{10}$, then there is an agreement with the observed value of the INS peak and the estimated $T_K$ from the peak position of the susceptibility for T=Ru and Os [38]. However, for T= Fe, $T_K$=210 K=18 meV is slightly higher than the INS peak at 12 meV. It is interesting to note that the high energy INS study reveals very broad magnetic excitations at 50 meV in $CeFe_2Al_{10}$ [46], while two broad CEF excitations have been observed in $CeT_2Al_{10}$ (T=Ru and Os) [61]. Further, it is interesting to compare the observed spin gap value in $CeT_2Al_{10}$ with that of the charge gap estimated through the optical spectroscopy [62-64]. The spectrum of polarized optical conductivity, $\sigma(T)$, along the b-axis differs greatly from that in the ac-plane, indicating that these materials have an anisotropic electronic structure. For $CeFe_2Al_{10}$ a charge gap of 55 meV was observed in all directions, but without a clear onset temperature, while for $CeRu_2Al_{10}$ and $CeOs_2Al_{10}$ a charge gap of 35 meV and 45 meV, respectively was observed in the a-c plane [63]. The observed excitations in the ac-plane originate from the optical transition across the *c-f* hybridization gap [62-64]. The observed peak structure at 20 meV was attributed to the formation of charge density waves (CDWs) in $CeRu_2Al_{10}$ and $CeOs_2Al_{10}$.

**2.2.2 INS study on single crystals of $CeT_2Al_{10}$**

We have seen a clear sign of spin gap formation in our INS study on the polycrystalline $CeT_2Al_{10}$ samples as discussed above. In order to further investigate the Q-dependence of these gapped excitations, we have carried out an INS study using the MERLIN spectrometer at the ISIS Facility, UK. Here we will present a short summary of our results and a detail study will be published elsewhere [61]. We performed omega scans between 0 and 90 deg with steps of two degrees for

CeT$_2$Al$_{10}$ (T=Ru and Os), whilst CeFe$_2$Al$_{10}$ was measured at fixed positions of the crystals, with the b- and c-axis // k$_i$, with respect to the incident beam. Multi crystals of CeFe$_2$Al$_{10}$ (total weight of 8.5g) and CeOs$_2$Al$_{10}$ (2.5g) were aligned with the b-c plane in the horizontal scattering plane (a-axis vertical), while the a-b plane was the horizontal scattering (c-axis vertical) for CeRu$_2$Al$_{10}$ (2.5g). We have used HORACE software [65] to combine all the data into a single S (Q, ω) file, which contains four dimensional data set, Q$_x$, Q$_y$, Q$_z$ and ω and hence the data can be projected out in any direction.

**(a) Spin wave investigations in CeT$_2$Al$_{10}$ with T=Ru and Os**

Figs.6 (a & b) show 2-dimensional slices of the data along the (1 -2 L) direction for CeRu$_2$Al$_{10}$ and the (1 0 L) direction for CeOs$_2$Al$_{10}$, respectively measured at 5 K. The magnetic excitations emerge from the antiferromagnetic zone centre with propagation vector k=(1 0 0). It is to be noted that the propagation vector k=(0 1 0) proposed for CeOs$_2$Al$_{10}$ from single crystal diffraction study is equivalent to (1 0 0) as they are connected through the reciprocal vector (-1 1 0). It is clear that we have observed well defined dispersive gapped magnetic excitations in both compounds. The gap near the zone centre is about 4-5 meV, whose origin could be due to a combined effect of the single ion anisotropy and anisotropic magnetic exchange. The anisotropic exchange could arise due to the presence of strong anisotropic hybridization between 4f-electrons and conduction electrons proposed from various experimental techniques, such as optical spectroscopy [62-64]. The zone boundary energy is 8 meV in CeRu$_2$Al$_{10}$ and 12 meV in CeOs$_2$Al$_{10}$. As both compounds exhibit a long-range antiferromagnetic magnetic ordering below 30 K, the observed magnetic excitations are mainly attributed to the spin waves from the antiferromagnetic ground state. Further, we have observed similar gapped excitations having similar values of the gap and zone boundary energies along the H- and K-directions. It is to be noted that the our TOF data of CeRu$_2$Al$_{10}$ is in agreement with that reported by J. Robert et al using the triple axis spectrometers, 2T at LLB, Saclay and IN8 and IN20 at ILL, Grenoble [66].

In order to gain insight into the exchange interaction in CeT$_2$Al$_{10}$ (T=Ru and Os, we have carried out extensive analysis of the spin wave spectrum using the McPhase program [67], in which spin waves (SW) are calculated by the mean field random phase approximation (MF-RPA). The Hamiltonian used in the SW calculation is given as follow:

$$H = \sum_{n,m} (B_n^m O_n^m) - \sum_{\langle i\ j\rangle\ \alpha,\beta} J_{ij}^{\alpha\beta} S_i^\alpha \cdot S_j^\beta \quad (1)$$

The first term, representing the single ion anisotropy, originates from the crystal field potential. Here $B_n^m$ (n=2,4 and m=0,2,4) are the crystal field parameters and $O_n^m$ are the Stevens operators. The values of the crystal field parameters were used from ref. [47]. The second term represents the two-ion magnetic exchange interactions (Fig.3). The summation <i j> in the second term is over pairs of spins with each pair counted once so that the $J$ constants are per spin pair with $\alpha$ and $\beta$ = x, y, z (or a, b, c). For isotropic exchange interactions we have $J_{ij}^{xx} = J_{ij}^{yy} = J_{ij}^{zz}$, while for anisotropic interactions we have taken $J_{ij}^{xx} = J_{ij}^{yy} \neq J_{ij}^{zz}$. First we tried to analyse our data using three isotropic exchanges interactions, however, we were unable to find any reasonable solution that reproduces our data satisfactorily. We therefore modified the first nearest neighbour (NN) interaction as anisotropic exchange and the remaining two NNN interactions as isotropic. From an extensive search over a wider parameter space we could find a solution that explains our data satisfactorily. The simulated data are presented in Fig.6 (c and d) and the exchange parameters (in meV) are: $J_{14}^{xx} = J_{14}^{yy} = -0.78$, $J_{14}^{zz} = -5.7$, $J_{12}^{xx} = J_{12}^{yy} = J_{12}^{zz} = 0.053$ and $J_{12}^{xx} = J_{12}^{yy} = J_{12}^{zz} = -0.178$ for CeOs$_2$Al$_{10}$ and : $J_{14}^{xx} = J_{14}^{yy} = -0.233$, $J_{14}^{zz} = -5.00$, $J_{12}^{xx} = J_{12}^{yy} = J_{12}^{zz} = 0.078$ and $J_{12}^{xx} = J_{12}^{yy} = J_{12}^{zz} = -0.095$ for CeRu$_2$Al$_{10}$. It is clear that estimated NN exchange interactions are highly anisotropic with $J^{zz} >> J^{xx}=J^{yy}$ in both T=Ru and Os compounds. The results of CeRu$_2$Al$_{10}$ are in agreement with that reported by Robert et al [66].

**(b) Hybridization gap study in the Kondo insulator CeFe$_2$Al$_{10}$**

We will briefly discuss the results of our INS study on CeFe$_2$Al$_{10}$, which does not exhibit any sign of a long-range magnetic ordering down to 50 mK. It is to be noted that as the ground state of CeFe$_2$Al$_{10}$ is paramagnetic, we do not expect any contribution from such a spin wave as we observed in CeRu$_2$Al$_{10}$ and CeOs$_2$Al$_{10}$ discussed previously. Hence CeFe$_2$Al$_{10}$ provides a unique opportunity to investigate pure contribution of spin gap or hybridization gap formation and its Q-and temperature dependence through INS study. We have seen well defined dispersive excitations in CeFe$_2$Al$_{10}$ for b//k$_i$ and c//k$_i$ directions at 5 K with a spin gap of ~8 meV for (0 0 L) and ~10 meV for (0 H 0) and (0 K 0) directions. Fig. 7a shows a color coded scattering intensity plot for (H 0 0) vs energy transfer obtained from c//k$_i$ at 5 K, which reveals that a column type scattering above 10 meV. It is interesting to note that despite having paramagnetic ground state of CeFe$_2$Al$_{10}$ the excitations emerge from a similar Q-position as observed in CeT$_2$Al$_{10}$ (T=Ru and Os), which have a long-range magnetic ground state. This indicates that the propagation vector for the magnetic excitations is also k=(1 0 0) in CeFe$_2$Al$_{10}$ (see Fig. 7c-e), in agreement with the single crystal INS study on CeFe$_2$Al$_{10}$ by Mignot et al [54]. It is to be noted that the excitations in CeFe$_2$Al$_{10}$ are somewhat different than the spin wave as it has intensity only for the branch going toward Q=0 from (1 0 0), but no or very weak intensity for the other branch going away from Q=0 i.e. going from (1 0 0) to (2 0 0) direction. Further, when the temperature was increased to 100 K (Fig.7b & Fig.7f-g) the excitations disappeared, which further confirms that the excitations are magnetic in nature and not associated with the lattice degrees of freedom. Overall, the results are in agreement with that obtained from the triple-axis study on CeFe$_2$Al$_{10}$ by Mignot et al [54], which shows a spin gap of 10 meV and zone boundary energy of 12 meV, with a very similar response in both H- and K-directions. However, we have seen clear anisotropy in the zone boundary energy between (H 0 0) and (0 K 0) and further the zone boundary energy along (H 0 0) direction is about 16-18meV or more, while it is less than this value for (0 K 0]) see Figs.7e & h.

We now compare the magnetic excitations in CeFe$_2$Al$_{10}$ with those observed in the mixed valence Kondo insulators SmB$_6$ and YbB$_{12}$ [68-70]. It is interesting to note that the various surface techniques, such as angle-resolved photoemission spectroscopy (ARPES) have shown that SmB$_6$ and YbB$_6$ exhibit a topological Kondo insulator (TKI) behaviour [71, 72]. The TKI behaviour in these compounds has been proposed based on theoretical models [73,74]. TKIs have a metallic surface state which is topologically protected by the bulk insulating state with an energy gap [73-74]. Recent neutron scattering studies on SmB$_6$ reveal a sharp inelastic peak centred near 14 meV, which is strongly Q and temperature dependent and the intensity nearly disappears above 60-80 K [68,69]. Further from the analysis of recent neutron time-of-flight INS data, including a third neighbour dominated hybridized band structure, it has been shown that SmB$_6$ is a TKI and exhibits a spin exciton below the charge gap [69]. The INS study of YbB$_{12}$ reveals two spin gap type excitations near 15.5 meV and 20 meV [70]. From the Q and temperature dependence it was found that the origin of 15.5 meV peak in YbB$_{12}$ is very similar to 14 meV peak in SmB$_6$. The 20 meV peak in YbB$_{12}$ was identified as manifesting a transition from a localized Kondo singlet with a gap to a magnetic state [70]. Further two spin gaps at 27 meV and 50 meV have been also observed through INS study on TKI CeOs$_4$Sb$_{12}$, which been interpreted as indirect (or spin gap) and direct (or charge gap) [75].

The origin of the magnetic excitations in CeFe$_2$Al$_{10}$ can be understood based on various theoretical models [6,19, 25,26, 76-77]. A very simple picture of Kondo insulators was proposed by Aeppli and Fisk based on two bands, and excitations associated with the transition from the lower band to upper band [76]. Riseborough et al have shown in more details that the spin-exciton observed in Kondo insulators can be derived as a magnetic excitation of the paramagnetic state of the Anderson Lattice Model by using the Random PhaseApproximation (RPA) [6,19, 77]. This model shows for a cubic system that the magnetic excitations arise from the antiferromagnetic zone centre. Further the intensity is maximum at the AFM zone centre and decreases strongly going away from it. Furthermore, with increasing $J_{cf}$ the spin-excitation dispersion softens and its intensity increases at the zone centre [77]. Mignot et al [54] have used a magnetic exciton model, originally proposed by Riseborough [77], to explain the observed magnetic excitations in $CeFe_2Al_{10}$. Another model was proposed by Ikeda and Miyake (IM) based on anisotropic hybridization in k-space to explain the observed properties of Kondo insulator CeNiSn [25]. The IM model corresponds to three symmetry related points in the space of crystal field ground states, where nodes develop. Moreno and Coleman (MC) have proposed a theoretical model to explain a gap-anisotropy in Kondo insulators, CeNiSn and CeRhSb [26]. Both IM and MC models involve a specific crystal field ground state for the Ce ions.

It is interesting to mention that mechanisms reminiscent of the spin-exciton formation, but based on the existence of a superconducting rather than insulating gap, have been invoked to explain anomalous magnetic modes occurring in high-Tc cuprates and Fe-based superconductors, the well-known ''resonance peak'' [79, 80, 81] or the HF superconductor compounds, $CeCu_2Si_2$ [82], $CeCoIn_5$ [83] and $UPd_2Al_3$ [84]. Interestingly to note that the spin resonance appears below Tc in the superconductors and its energy is related to $T_c$ with the well know relation $\Delta_{reso}$~4.3*$T_c$ (for pnictides) and ~5.4*$T_c$ (for cuprates) [85, 86], while the spin-gap in Kondo insulators appears at a temperature below $T^\chi_{max}$ and its value is related to $T_K$ with relation $\Delta_{spin}$~3* $T^\chi_{max} = T_K$ [28]. Further a similar resonant mode in the non-superconducting antiferromagnetic heavy-fermion metal $CeB_6$ has been observed at the Q-position corresponding to antiferro-quadrupolar (AFQ) ordering (½, ½, ½) [87]. Akbari and Thalmeier [88] have proposed a theoretical model based on a fourfold degenerate Anderson lattice model, which explains the appearance of the spin exciton resonance and the momentum dependence of its spectral weight, in particular, around the AFQ vector and its rapid disappearance in the disordered phase of $CeB_6$. The pictures emerging here for an archetype KI may thus prove of interest for a broader class of systems exhibiting spin gap or spin resonance.

## 3. Conclusions

We have carried out a suite of comprehensive inelastic neutron scattering measurements on Kondo insulators $CeT_2Al_{10}$ (T=Fe, Ru and Os). We have seen a clear sign of a spin gap formation at low temperature through inelastic neutron scattering in T=Fe, Ru and Os polycrystalline samples. The energy of the spin gap seems related to the hybridization strength or Kondo temperature estimated from the observed maximum in the static susceptibility. Further, the temperature dependent INS study clearly indicates that the spin gap does exist well above $T_N$ in both T=Ru and Os compounds, which may indicate that the origin of the spin gap is not purely due to a gapped spin wave, but is also associated with the hybridization gap. The INS study on the single crystals of T=Fe, Ru and Os has established the formation of a spin energy-gap in all three systems, despite that T=Fe compound does not order magnetically down to 50 mK. Analysis of the observed magnetic/spin-wave excitations in T=Ru and Os reveals the presence of strong anisotropic magnetic exchange, $J^{xx}=J^{yy}<<J^{zz}$, explaining the anomalous direction of the ordered state moment, which is along the c-axis and not along the a-axis as expected from the strong single ion CEF anisotropy. We believe the coexistence of the Kondo semiconducting state with spin-gap formation and magnetic order in $CeT_2Al_{10}$ (T=Ru, Os) to be unique among 4*f*-electron systems, so far not observed in many systems, and it poses a perplexing new ground state for the strongly correlated class of materials. This makes $CeT_2Al_{10}$ a particularly attractive model system to study the interplay between RKKY interaction, Kondo screening and quantum criticality in the global phase diagram of heavy fermion metals.

**Acknowledgement:**

We acknowledge interesting discussions with Profs. Andre Strydom, Peter Riseborough, Qimiao Si, Piers Coleman, Pavel Alekseev and Drs Jean-Michel Mignot, Julian Robert, Louis-Pierre Regnault and Tatiana Guidi. DTA would like to thank CMPC-STFC, grant number CMPC-09108 for financial support, and JSPS for the award of their fellowship, which enabled him to visit Hiroshima University during this period this manuscript was written. The work at Hiroshima University and Toyama Prefectural University was supported by JPSJ KAKENHI Grant Nos. 26400363, 15K05180 and 16H01076.

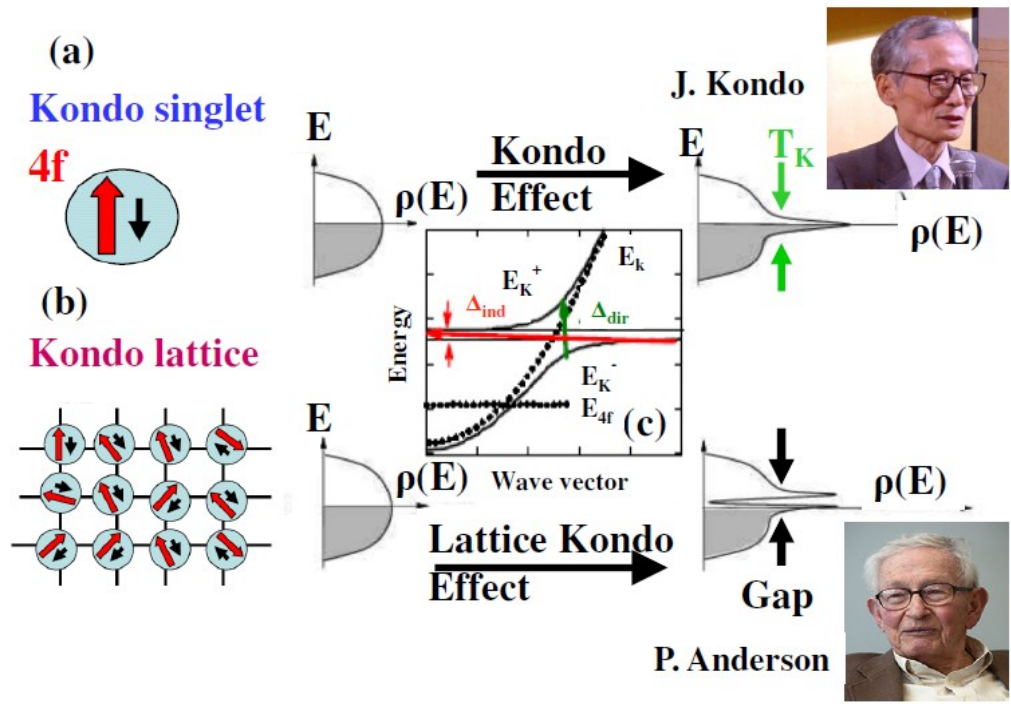

Fig. 1 (color online) (a) (left) Single impurity Kondo effect, coupling between local 4f moment (big red arrow) and conduction electrons, $J_{cf}$ (small black arrow), (middle) a schematic view of the conduction electron density of states (DOS) with no Kondo effect (i.e. $J_{cf}$=0), (right) a schematic view of DOS with presence of Kondo effect, which shows a build-up of the fermionic resonance near the Fermi level. The width of the resonance peak gives an estimation of Kondo temperature, $T_K$. (b) (right) Kondo lattice effect in real space, (middle) a schematic view of DOS with no Kondo effect, (right) a schematic view of DOS with presence of Kondo lattice effect, which shows an opening of a small energy gap in the fermionic resonance near $E_F$. This class of materials with a small gap near $E_F$ is known as Kondo insulators or Kondo semimetals. (c) Kondo lattice effect in k-space, hybridized band picture showing renormalized bands, lower ($E_K^-$) and upper ($E_K^+$) hybridized bands, and direct gap, $\Delta_{dir}$ (we called charged gap, $\Delta_{char}$, shown by a vertical green arrow) at q = 0 and indirect gap, $\Delta_{ind}$ (we called spin gap, $\Delta_{spin}$, shown by a horizontal red arrow) at q ≠ 0 and the gap in resonant density of states (DOS). The dotted line shows unhybridized conduction band with a parabolic dispersion and localised 4f band (after refs. [18,19]).

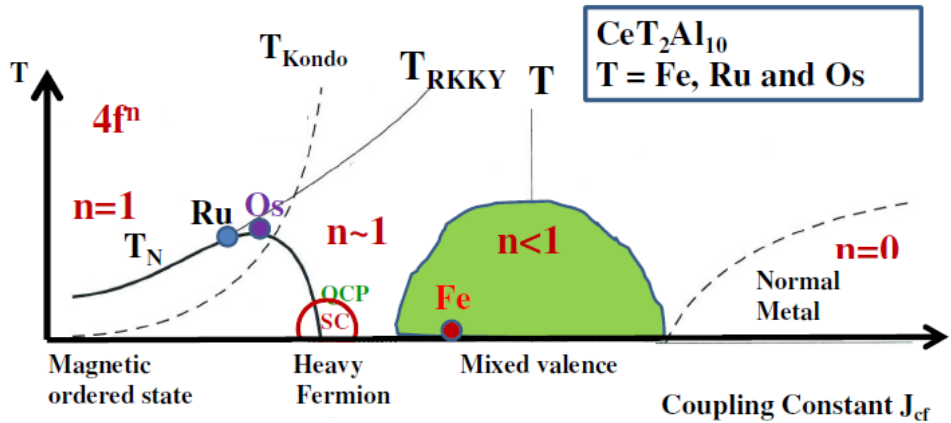

Fig.2 (color online) Schematic view of the Doniach phase diagram (after D.I. Khomskii. [89], plotted as characteristic temperature versus strength of coupling constant $J_{cf}$ between 4f moment and conductions. When magnetic ordering temperature $T_N$ vanishes to zero at the critical coupling, new regime appears (called quantum critical point, QCP) where quantum fluctuations dominate over thermal fluctuations. In this new regime unconventional superconductivity (SC) has been observed. The three filled circles represent a relative position of $CeT_2Al_{10}$ (T=Fe, Ru and Os) on the phase diagram.

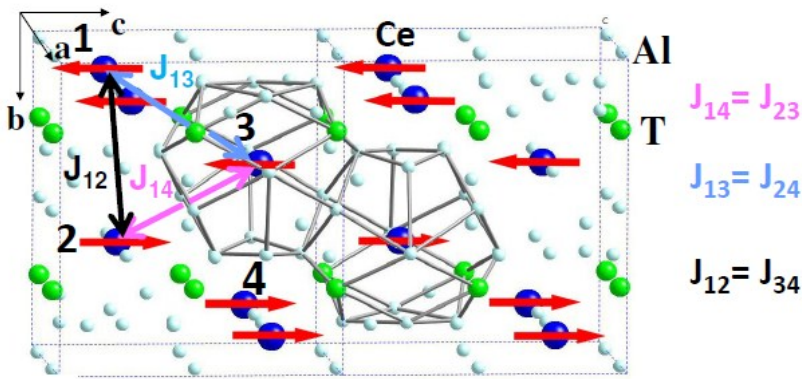

Fig.3 (color online) The caged type orthorhombic unit cell of $CeT_2Al_{10}$ (T=Fe, Ru and Os) and magnetic structure of $CeRu_2Al_{10}$ from ref. [42]. The various exchange paths are shown by a purple, light blue and black arrow.

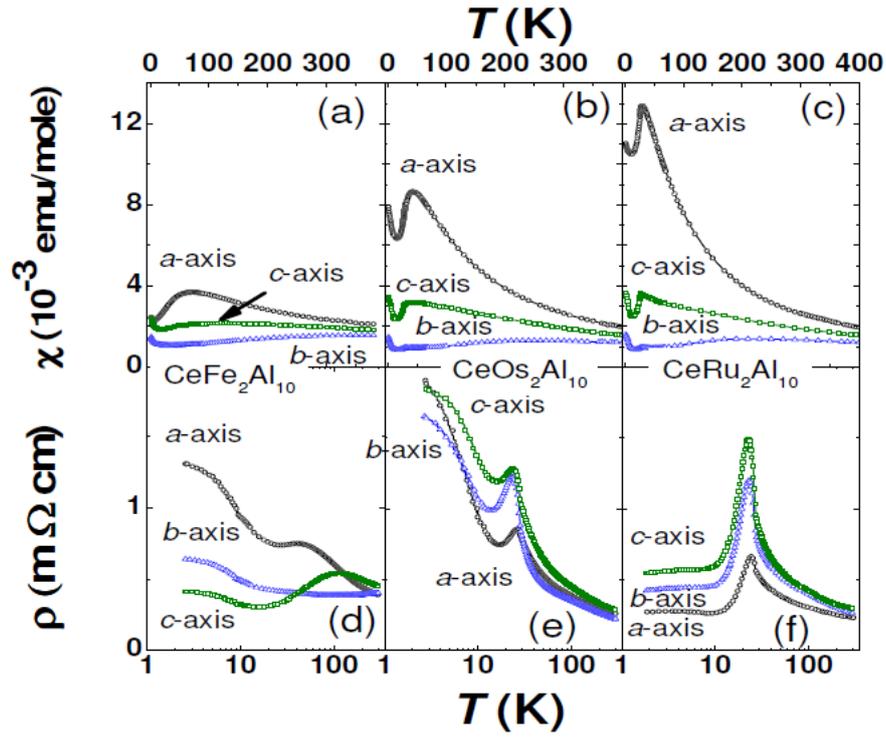

Fig.4 (a-c) Temperature dependence of the magnetic susceptibility of $CeT_2Al_{10}$ (T=Fe, Ru and Os) and (d-f) temperature dependence of the electrical resistivity of $CeT_2Al_{10}$ (T=Fe, Ru and Os) from ref. [28]

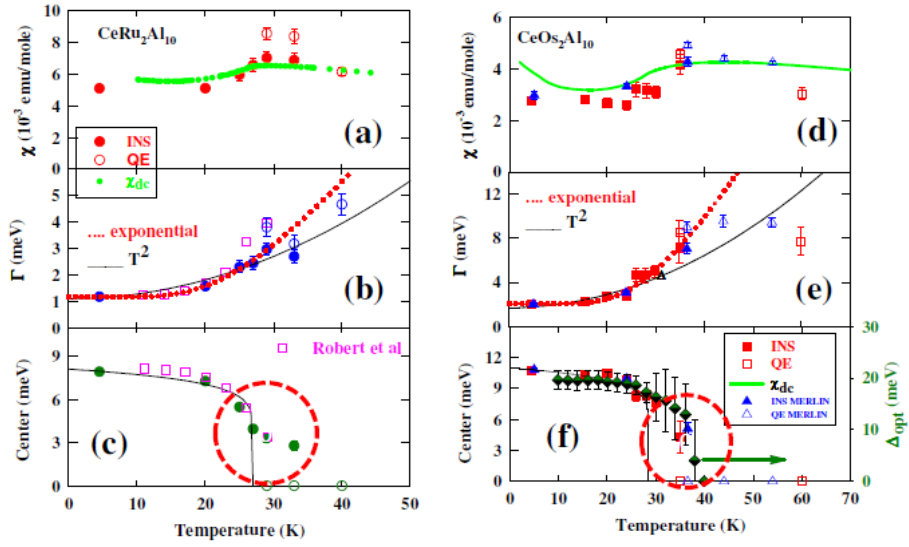

Fig.5 (color online) The fit parameters, susceptibility, linewidth and peak position versus temperature obtained from fitting the magnetic inelastic scattering intensity of $CeRu_2Al_{10}$ (left) and $CeOs_2Al_{10}$ (right) (detail can be found in refs. [28,46]). The closed circles represent the fit using an inelastic peak and open circles represent the fit using a quasi-elastic peak. The green solid line in (top figs) shows the measured dc-susceptibility of the polycrystalline sample of $CeRu_2Al_{10}$ (a), and $CeOs_2Al_{10}$. The middle figures show the linewidth versus temperature and the dotted and solid line represents the fits using an exponential behaviour and $T^2$ behaviour respectively (see text). The bottom figures show the centre of inelastic peak (or $\Delta_{spin}$) versus temperature. The filled green diamonds in (f, right y-axis) are the optical gap ($\Delta_{op}$) from ref. [62]. The solid line represents the power law simulation $\Delta(T) = \Delta_0 (1-T/T_N)^\beta$ with $\Delta_0 = 8.1$ meV, $\beta = 0.1$ and $T_N = 27$ K for $CeRu_2Al_{10}$ from ref. [46] and $\Delta_0 = 11$ meV, $\beta = 0.1$ and $T_N = 28.5$ K for $CeOs_2Al_{10}$ from ref. [28].

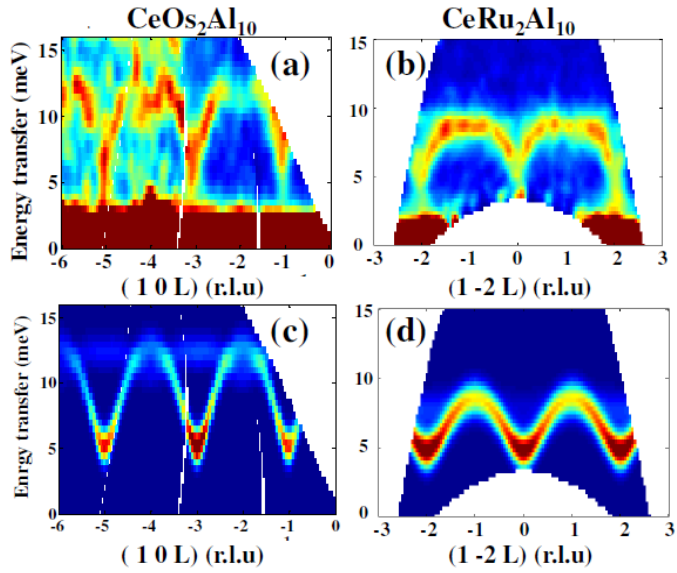

fig. 6 (color online) Color coded inelastic scattering (or spin wave) intensity maps of $CeOs_2Al_{10}$ (a) and $CeRu_2Al_{10}$ (b) at 5 K. The data were collected using HORACE or omega scans from 0 to 90 degrees in steps of 2 degree and using an incident neutron energy of $E_i = 25$ meV. Bottom figures show the simulated spin wave excitations using the McPhase program (ref,[67]) for $CeOs_2Al_{10}$ (c) and $CeRu_2Al_{10}$ (d) (see text).

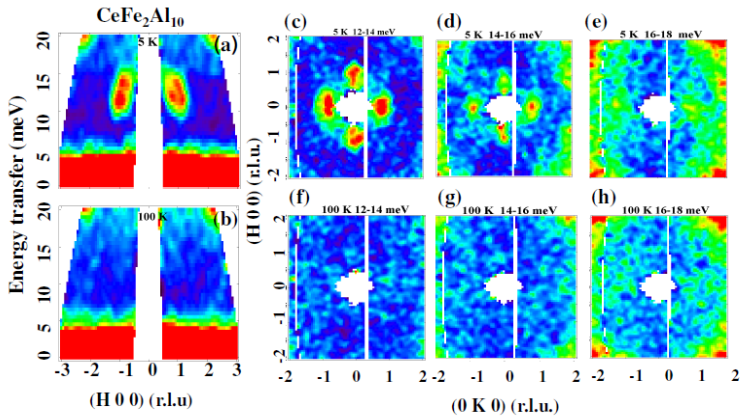

fig. 7 (color online) Color coded inelastic scattering (or spin gap excitations) intensity maps of $CeFe_2Al_{10}$ at 5 K (a) and at 100 K (b). The data were collected with fixed position of crystal, b-axis // $k_i$ using an incident neutron energy of $E_i = 40$ meV. Figs. (c-e) and Figs. (f-h) show a constant energy slices in the a-b plane with increasing energy, at 5 K and 100 K, respectively. 12-14 meV (c & f), 14-16 meV (d & g) and 16-18 meV (e & h).

**References:**

[1] H. v. Löhneysen, A. Rosch, M. Vojta, and P. Wölfle, Rev. Mod. Phys. **79**, 1015 (2007); F. Steglich, Mag. Mag. Matt. **100**, 186 (1991)
[2] C.M. Varma, Rev. Mod. Phys. **48**, 219 (1976); J. M. Lawrence, P. S. Riseborough and R. D. Parks, Rep. Prog. Phys. **44**, 1 (1981)
[3] E. Bauer, Adv. Phys. **40**, 417 (1991)
[4] S K Dhar, S K Malik and R Vijayaraghavan, J. Phys. C: Solid State Phys., **14,** L321 (1981)
[5] F. Steglich et al., Phys. Rev. Lett. **43**, 1892 (1979); V.P. Mineev and K.V. Samokhin, Introduction to Unconventional Superconductivity (Gordon and Breach, London, 1999)
[6] P.S. Riseborough, Adv. Phys. **49**, 257 (2000)
[7] J. Spałek and A.Slebarski**,** J. Physics: Conf. Series **273,** 012055 (2011)
[8] A. Georges, G. Kotliar, W. Krauth and M.J. Rozenberg, Rev. Mod. Phys. **68**, 13 (1996)
[9] G.R. Stewart, Rev. Mode. Phys. **56**, 755 (1984); J. Flouquet and H. Harima, arXiv:0910.3110 (2009)
[10] Q. Si, Phys. Status Solidi B **247**, 476 (2010)
[11] P. Coleman and A.J. Schofield, Nature **433**, 226-229 (2005)
[12] K. Andres, J.E. Graebner and H.R. Ott, Phys. Rev. Lett. **35**, 1779 (1975)
[13] J.W. Allen, B. Batlogg and P, Wachter, Phys. Rev. B **20**, 4807 (1979)
[14] P.W. Anderson, Phys. Rev. **124**, 41 (1961)
[15] J. Kondo, in Solid State Physics, edited by H. Ehrenreich and D. Turnbull (Academic Press, New York, 1964), Vol. **23**, p. 183; J. Kondo, Progr. Theor. Phys. **32**, 37 (1964)
[16] J. R. Schrieffer and P. Woll, Phys. Rev. **149**, 491 (1966)
[17] M.A. Ruderman and C. Kittel, Phys. Rev. **96**, 99 (1954); T. Kasuya, Prog. Theor. Phys. **16**, 45 (1956); K. Yosida, Phys. Rev. **106**, 893 (1957)
[18] P. Coleman, "Introduction to Many-Body Physics", (Cambridge University Press, 2015); P. Coleman, Handbook of Magnetism and Advanced Magnetic Materials Vol. 1, (Wiley, New York, 2007)
[19] P.S. Riseborough, Phys. Rev. B **45**, 13984 (1992)
[20] T. Takabatake, F. Teshima, H. Fujii, S. Nisjhigori, T. Suzuki, T. Fujita, Y. Yamaguchi, J. Sakurai, and D. Jaccard, Phys. Rev. B **41**, 9607 (1990); H. Kadowaki et al., J. Phys. Soc. Jpn. **63**, 2074 (1994)
[21] S.K. Malik and D.T. Adroja, Phys. Rev. B **43**, 6277 (1991)
[22] B. Bucher, Z. Schlesinger, P.C. Canfield, Z. Fisk, Physica B **199&200**, 489 (1994).
[23] P. A. Alekseev, J.-M. Mignot, J. Rossat-Mignod, V. N. Lazukov, I. P. Sadikov, E. S. Konovalova and Yu. B. Paderno, J. Phys.: Candens. Mattcr **7**, 289 (1995); E.E. Vainshtein, S.M. Blokhin and Yu.B. Paderno, Sov. Phys.–Solid State (Engl. Transl.), **6**, 2318 (1965)
[24] E. V. Nefeodova and P. A. Alekseev, J.-M. Mignot, V. N. Lazukov, I. P. Sadikov, Yu. B. Paderno, N. Yu. Shitsevalova, R. S. Eccleston, Phys. Rev. B **60**, 13507 (1999); F. Iga, S. Hiura, J. Klijn, N. Shimizu, T. Takabatake, M. Ito, F. Masaki, Y. Matsumoto, T. Suzuki, and T. Fujita, Physica B **259–261**, 312 (1999); M. Kasaya, F. Iga, K. Negishi, S. Nakai, and T. Kasuya, J. Magn. Magn. Mater. **31–34**, 437 (1983)
[25] H. Ikeda, K. Miyake, J. Phys. Soc. Jpn. **65**, 1769 (1996)
[26] J. Moreno and P. Coleman, Phys. Rev. Lett. **84**, 342 (2000)
[27] S. Doniach, Physica B **91**, 231 (1977); C. Lacroix and M. Cyrot, Phys. Rev. B **20**, 1969 (1979); R. Jullien, J. N. Fields, and S. Doniach Phys. Rev. B **16**, 4889 (1977); R. Jullien, J. Fields, and S. Doniach Phys. Rev. Lett. **38**, 1500 (1977).
[28] see for example, D T Adroja, A D Hillier, Y Muro, T Takabatake, A M Strydom, A Bhattacharyya, A Daoud-Aladin and J W Taylor, Physics Script, **88**, 068505 (2013)
[29] A.M. Strydom, Physica B **404**, 2981 (2009)

[30] Y. Muro. J. Kajino, K. Umeo, K. Nishimoto and R. Tamura and T. Takabatake, Phys. Rev. B **81**, 214401(2010**)**
[31]  K. Yutani, Y. Muro, J. Kajino, T. J. Sato and T. Takabatake, J. Phys. Conf. Ser. **391**, 012070 (2012)
[32] Y. Muro, K. Motoya, Y. Saiga and T. Takabatake, J. Phys.: Conf. Ser. **200**, 012136 (2010)
[33] Y. Muro, K. Motoya, Y. Saiga, and T. Takabatake, J. Phys. Soc. Jpn. **78**, 083707 (2009); T. Nishioka, Y. Kawamura, T. Takesaka, R. Kobayashi, H. Kato, M. Matsumura, K. Kodama, K. Matsubayashi, and Y. Uwatoko, J. Phys. Soc. Jpn. **78**, 123705 (2009)
[34] M. Matsumura, Y. Kawamura, S. Edamoto, T. Takesaka, H. Kato, T. Nishioka, Y. Tokunaga, S. Kambe, and H. Yasuoka, J. Phys. Soc. Jpn. **78** 123713  (2009)
[35] T. Nishioka, Y. Kawamura, T. Takesaka, R. Kobayashi, H. Kato, M. Matsumura, K. Kodama, K. Matsubayashi, and Y. Uwatoko: J. Phys. Soc. Jpn. **78**  123705 (2009)
[36] H. Tanida, D. Tanaka, M. Sera, C. Moriyoshi, Y. Kuroiwa, T. Nishioka, H. Kato, and M. Matsumura, J. Phys. Soc. Jpn. **79**, 043708 (2010); K. Yoshida R. Okubo, H. Tanida, T. Matsumura, M. Sera, T. Nishioka, M. Matsumura, C. Moriyoshi, and Y. Kuroiwa, Phys. Rev. B **91**, 235124 (2015)
[37] D. Khalyavin, A. D. Hillier, D. T. Adroja, A. M. Strydom, P. Manuel, L. C. Chapon, P. Peratheepan, K. Knight, P. Deen, C. Ritter, Y. Muro, and T. Takabatake, Phys. Rev. B **82**, 100405(R) (2010)
[38] D. T. Adroja, A. D. Hillier, P. P. Deen, A. M. Strydom, Y. Muro, J. Kajino, W. A. Kockelmann, T. Takabatake, V. K. Anand, J. R. Stewart, and J. Taylor, Phys. Rev. B **82** 104055 (2010)
[39]  H. Kato, R. Kobayashi, T. Takesaka, T. Nishioka, M. Matsumura, K. Kaneko, and N. Metoki, J. Phys. Soc. Jpn, Suppl. **80**, 073701 (2011).
[40] D. D. Khalyavin, D. T. Adroja,  P. Manuel, J. Kawabata, K. Umeo, T. Takabatake, and A. M. Strydom, Phys. Rev. B **88**, 060403(R) (2013**)**
[41] A. Bhattacharyya, D. D. Khalyavin,  D. T. Adroja, A. M. Strydom, A. D. Hillier, P. Manuel, T. Takabatake, J. W. Taylor and C. Ritter, Phys. Rev. B **90**, 174412 (2014);
[42] D. D. Khalyavin, D. T. Adroja, A. Bhattacharyya, A. D. Hillier,  P. Manuel, A. M. Strydom, J. Kawabata, and T. Takabatake, Phys. Rev. B **89**, 064422 (2014)
[43] A. Bhattacharyya, D. T. Adroja, A. M. Strydom, J. Kawabata, T. Takabatake, A. D. Hillier, V. Garcia Sakai, J. W. Taylor, and R. I. Smith, Phys. Rev. B  **90**, 174422 (2014)
[44] J. Robert, J.M. Mignot, G. André, T. Nishioka, R. Kobayashi, M. Matsumura, H. Tanida, D. Tanaka and M. Sera,  Phys. Rev. B **82** 100404 R (2010)
[45] J. M. Mignot, J. Robert, G. Andre, A. M. Bataille, T. Nishioka, R. Kobayashi, M. Matsumura, H. Tanida, D. Tanaka, and M. Sera, J. Phys. Soc. Jpn. **80,** SA022 (2011)
[46] D. T. Adroja, A. D. Hillier, Y. Muro, J. Kajino, T. Takabatake, P. Peratheepan, A. M. Strydom, P. P. Deen, F. Demmel, J. R. Stewart, J.W. Taylor, R. I. Smith, S. Ramos, and M. A. Adams, Phys. Rev. B **87**, 224415 (2013).
[47] F. Strigari, T. Willers, Y. Muro, K. Yutani, T. Takabatake, Z. Hu, Y.-Y. Chin, S. Agrestini, H.-J. Lin, C. T. Chen, A. Tanaka,M. W. Haverkort, L.-H. Tjeng, and A. Severing, Phys. Rev. B **86**, 081105(R) (2012)
[48] F. Strigari, T. Willers, Y. Muro, K. Yutani, T. Takabatake, Z. Hu, S. Agrestini, C.-Y. Kuo, Y.-Y. Chin, H.-J. Lin, T. W. Pi, C. T. Chen, E. Weschke, E. Schierle, A. Tanaka, M. W. Haverkort, L. H. Tjeng, and A. Severing, Phys. Rev. B **87**, 125119 (2013)
[49]  A. Kondo, K. Kindo, K. Kunimori, H. Nohara, H. Tanida, M. Sera, R. Kobayashi, T. Nishioka, and M. Matsumura, J. Phys. Soc. Jpn. **82**, 054709 (2013).;A. Kondo, J. Wang, K. Kindo, T. Takesaka, Y. Kawwamura, T. Nishioka, D. Tanaka, H. Tanida, and M. Sera, J. Phys. Soc. Jpn. **79**, 073709 (2010); H. Guo, H. Tanida, R. Kobayashi, I. Kawasaki, M. Sera, T. Nishioka, M. Matsumura, I. Watanabe, and Zhu-an Xu, Phys. Rev. B  **88**, 115206 (2013)
[50] G.L. Squires, Introduction to the theory of thermal neutron scattering, (Cambridge University Press, 1978)

[51] S.W. Lovesey, Theory of neutron scattering from condensed matter, (Clarendon Press, Oxford, 1984)
[52] W.G. Williams, Polarized Neutrons, (Clarendon Press, Oxford, 1988).
[53] S. M. Hayden, H. A. Mook, V. Dai, T. G. Perring, and F. Dogan, Nature (London) **429**, 531 (2004); R. Coldea, S. M. Hayden, G. Aeppli, T. G. Perring, C. D. Frost, T. E. Mason, S.-W. Cheong, and Z. Fisk, Phys. Rev. Lett. **86**, 5377 (2001)
[54] J.-M. Mignot, P. A. Alekseev, J. Robert, S. Petit, T. Nishioka, M. Matsumura, R. Kobayashi, H. Tanida, H. Nohara, and M. Sera, Phys. Rev. B **89**, 161103(R) (2014)
[55] A.J. Fedro and S.K. Sinha, in Valence Fluctuations in Solids, edited by L.M. Falicov, W. Hanke, and M.B. Maple (North-Holland, Amsterdam, p. 329, 1981)
[56] D. T. Adroja, K.A. McEwen. J.-G. Park, A.D. Hillier, N. Takeda, P.S. Riseborough and T. Takabatake, J. Opto. Adv. Mater., **10**, 1564 (2008)
[57] R.Viennois, L. Girard, L.C. Chapon, D.T. Adroja, R.I. Bewley, D. Ravot, P. S. Riseborough, and S. Paschen, Phys. Rev. B **76**, 174438 (2007)
[58] A. Severing, J. D. Thompson, P. C. Canfield, Z. Fisk and P. Riseborough, Phys. Rev. B **46**, 6832 (1991)
[59] T.J. Sato, H. Kadowaki, H. Yoshizawa, T. Ekino, T. Takabataket, H. Fujii, L.P. Regnault and Y. Isikawa, J. Phys.: Condens. Marter **7**, 8009 (1995); T.E. Mason, G. Aeppli, A.P. Ramirez, K.N. Clausen, C. Broholm, N. Stücheli, E. Bucher, and T.T.M. Palstra, Phys. Rev. Lett. **69** 490 (1992)
[60] D. T. Adroja, J.-G. Park, K. A. McEwen, N. Takeda, M. Isikawa, and J.-Y. So, Phys. Rev. B **68**, 094425 (2003)
[61] D.T. Adroja et al to be published
[62] S. Kimura, T. Iizuka, H. Miyazaki A. Irizawa, Y. Muro, and T. Takabatake, Phys. Rev. Lett. **106,** 056404 (2011)
[63] S. Kimura,T. Iizuka, H. Miyazaki,T. Hajiri, M. Matsunami, T. Mori, A. Irizawa, Y. Muro J. Kajino, and T. Takabatake, Phys. Rev. B **84**, 165125 (2011)
[64] S. Kimura, Y. Muro, and T. Takabatake, J. Phys. Soc. Jpn. **80**, 033702 (2011); S. Kimura, T. Iizuka, Y. Muro, J. Kajino and T. Takabatake,J. Phys. Conf. Series **391**, 012030 (2012)
[65] http://arxiv.org/abs/1604.05895
[66] J. Robert, J-M. Mignot, S. Petit, P. Steffens, T. Nishioka, R. Kobayashi, M. Matsumura, H. Tanida, D. Tanaka, and M. Sera, Phys. Rev. Lett. **109**, 267208 (2012)
[67] M. Rotter, M. D. Le, A. T. Boothroyd and J. A. Blanco, J. Phys. Cond Mat.: 24 (2012) 213201
[68] P.A. Alekscev,A.S. Ivanov, V.N. Lazukov, I.P. Sadikov and A. Severing, Physicn B **180-181** 281 (1992);P A Alekseev, J M Mignot, J Rossat-Mignod, V N Lazukov, I P Sadikov, E S Konovalova and Yu B Paderno, J. Phys. Cond. Matte. **7**, 289-305 (1995); P.A. Alekseev, J.-M. Mimot, J. Rossat-Mignod V.N. Lazukov and I.P. Sadikov, Phwicu B **186-188**, 334 (1993)
[69] W. T. Fuhrman, J. Leiner, P. Nikolić, G. E. Granroth, M. B. Stone, M. D. Lumsden, L. DeBeer-Schmitt, P. A. Alekseev, J.-M. Mignot, S. M. Koohpayeh, P. Cottingham, W. Adam Phelan, L. Schoop, T. M. McQueen, and C. Broholm, Phys. Rev. Lett. **114**, 036401 (2015)
[70] K. S. Nemkovski, J.-M. Mignot, P. A. Alekseev, A. S. Ivanov, E. V. Nefeodova, A. V. Rybina, L.-P. Regnault, F. Iga, and T. Takabatake, Phys. Rev. Lett. **99**, 137204 (2007); J.-M. Mignot, P. A. Alekseev, K. S. Nemkovski, L.-P. Regnault, F. Iga, and T. Takabatake, Phys. Rev. Lett. **94**, 247204 (2005)
[71] F. Lu, J. Zhao, H. Weng, Z. Fang, and X. Dai, Phys. Rev. Lett. **110**, 096401 (2013)
[72] H. Weng, J. Zhao, Z. Wang, Z. Fang, and X. Dai, Phys. Rev. Lett. **112**, 016403 (2014)
[73] V. Alexandrov, M. Dzero, and P. Coleman, Phys. Rev. Lett. **111**, 226403 (2013)
[74] M. Dzero, K. Sun, V. Galitski, and P. Coleman, Phys. Rev. Lett. **104**, 106408 (2010); T. Takimoto, J. Phys. Soc. Jpn. **80**, 123710 (2011)
[75] D.T. Adroja, J.-G. Park, E. A. Goremychkin, K. A. McEwen, N. Takeda, B. D. Rainford, K. S. Knight, J. W. Taylor, Jeongmi Park, H. C. Walker, R. Osborn, and Peter S. Riseborough , Phys. Rev. B **75**, 014418 (2007); B. Yan, L. Müchler, X.-L. Qi, S.-C. Zhang and C. Felser, Phys. Rev. B **85**, 165125 (2012)

[76] G. Aeppli and Z. Fisk, Comment Cond. Mater Phys. **16**,155 (1992)
[77] P. S. Riseborough, J. Magn. Magn. Mater. **226–230**, 127 (2001);P.S. Riseborough, Ann. Phys.(Leipzing), **9**, 813 (2000) ;P.S. Riseborough and S.G. Magalhaes, J. Magn. Magn. Mater, **400**, 3 (2016)
[79] J. Rossat-Mignod, L.P. Regnault, C. Vettier , P. Bourges , P. Burlet, J. Bossy, J Y. Henry and G. Lapertot, Physica (Amsterdam) **185–189C**, 86 (1991)
[80] F. Onufrieva and P. Pfeuty, Phys. Rev. B **65**, 054515 (2002)
[81] A. D. Christianson, E. A. Goremychkin, R. Osborn, S. Rosenkranz, M. D. Lumsden, C. D. Malliakas, I. S. Todorov, H. Claus, D. Y. Chung, M. G. Kanatzidis, R. I. Bewley, and T. Guidi, Nature **456**, 930 (2008).
[82] O. Stockert, J. Arndta, A. Schneidewind, H. Schneider, H.S. Jeevan. C. Geibel, F. Steglich, M. Loewenhaupt, Physica B **403**, 973 (2008); .I. Eremin, G. Zwicknagl, P. Thalmeier, and P. Fulde, Phys. Rev. Lett. **101**, 187001 (2008)
[83] C. Stock, C. Broholm, J. Hudis, H. J. Kang, and C. Petrovic, Phys. Rev. Lett. **100**, 087001 (2008)
[84] N. K. Sato et al., Nature (London) **410**, 340 (2001)
[85] C. Zhang, W. Lv, G. Tan, Yu. Song, S. V. Carr, S. Chi, M. Matsuda, A. D. Christianson, J. A. Fernandez-Baca, L. W. Harriger, and Pengcheng Dai, arXiv:1605.06890; Phys. Rev. B in press (2016)
[86] D. S. Inosov, J. T. Park, A. Charnukha, Yuan Li, A. V. Boris, B. Keimer, and V. Hinkov, Phys. Rev. B **83**, 214520 (2011); G. Yu, Y. Li, E. M. Motoyama, and M. Greven, Nat. Phys. **5**, 873 (2009)
[87] G. Friemel , Y. Li, A.V. Dukhnenko, N.Y. Shitsevalova, N.E. Sluchanko, A. Ivanov, V.B. Filipov, B. Keime, and D.S. Inoso, Nat. Comm. **3**, 830 (2012)
[88] A. Akbari and P. Thalmeier, Phys. Rev. Lett. **108**, 146403 (2012)
[89] D.I. Khomskii, Basnic Aspects of the Quantum Theory of Solids, Order and Elementary Excitations, (Cambridge University Press, 2010)